\documentclass{article}
\usepackage[utf8]{inputenc}

\title{An investigation of pre-upsampling generative modelling and Generative adversarial networks in Audio Super resolution\footnotemark}
\usepackage{tablefootnote}
\author{James King, Ramon Viñas Torné, Alexander Campbell, Pietro Liò}
\date{July 2021}
\usepackage{hyperref}
\usepackage{url}
\usepackage[numbers]{natbib}
\usepackage{graphicx}
\usepackage{ICLRsyle,times}
\usepackage{tabularx}
\usepackage{booktabs}
\usepackage{wrapfig}
\usepackage{float}
\usepackage{amsmath}
\usepackage{amsfonts}
\usepackage{amssymb}
\usepackage{gensymb}
\usepackage{enumitem}
\usepackage{multirow}
\usepackage{cleveref}
\usepackage{siunitx}
\usepackage{makecell}
\usepackage{algorithmic}
\usepackage{algorithm}

\iclrfinaltrue
\setcitestyle{number}
\begin{document}
\maketitle
\begin{abstract}
\footnotetext[1]{Supported by University of Cambridge.}
There have been several successful deep learning models that perform audio super-resolution. Many of these approaches involve using preprocessed feature extraction which requires a lot of domain-specific signal processing knowledge to implement. Convolutional Neural Networks (CNNs) improved upon this framework by automatically learning filters. An example of a convolutional approach is AudioUNet, which takes inspiration from novel methods of upsampling images. Our paper compares the pre-upsampling AudioUNet to a new generative model that upsamples the signal \textit{before} using deep learning to transform it into a more believable signal. Based on the EDSR network for image super-resolution, the newly proposed model outperforms UNet with a 20\% increase in log spectral distance and a mean opinion score of 4.06 compared to 3.82 for the two times upsampling case. AudioEDSR also has 87\% fewer parameters than AudioUNet. How incorporating AudioUNet into a Wasserstein GAN (with gradient penalty) (WGAN-GP) structure can affect training is also explored. Finally the effects artifacting has on the current state of the art is analysed and solutions to this problem are proposed. The methods used in this paper have broad applications to telephony, audio recognition and audio generation tasks.
\end{abstract}
\section{Introduction}
\textbf{Audio Super-resolution} is the process of artificially increasing the sample rate of a signal. It is a common approach to performing \textbf{bandwidth extension}, which expands the frequency range of a signal. These terms are often used interchangeably in literature, and the Nyquist-Shannon Theorem \cite{1697831} shows that sample rate and signal frequencies are proportional to each other. There are several applications of this type of super-resolution work. It can be used in speaker verification networks, improving the quality of telecommunications, audio style transfer networks, and any system which generates audio signals. 

This paper looks at methods that perform operations directly on raw audio signals. These operations are expensive and require a large number of flops per second. To address these issues, the parallelism property of fully convolutional architectures can be exploited. Several modern deep learning approaches to solving the problem of audio super-resolution are analysed on how they differ in terms of size and effectiveness. The best ways of training these models are also evaluated. Next a new system based on a Generative Adversarial Network (GAN) is introduced and the effect this has on training is explored. Furthermore, the problem of artifacting is studied and new solutions for future research are proposed.

\section{Setup and Background}
An audio signal can be represented as a function of time $w(t):[0,L]\rightarrow\mathbb{R}$ where $L$ is the duration of the signal, and $w(t)$ is the amplitude of the signal at time $t$. In order to operate on this data computationally $w(t)$ needs to be discretised into $\Tilde{w}(t;S):\{\frac{1}{S},\frac{2}{S},\hdots,\frac{SL}{S}\}\rightarrow\mathbb{R}$ where $S$ is the sampling rate or \textbf{resolution} of the discretised signal $\Tilde{w}$. Suppose the same signal is discretised with two different resolutions $u = \Tilde{w}(t;S)$ and $l = \Tilde{w}(t;\hat{S})$ with $S > \hat{S}$. Our models aim to learn over $\mathbb{P}(u|l)$ by assuming that $u = f_{\theta}(l)+\vec{\epsilon}$ for a model $f_{\theta}$ with parameters $\theta$ and $\vec{\epsilon}\sim\mathcal{N}(0,1)$.

We aim to develop more light weight models than the current state of the art and in particular our models are trained to upsample to 12kHz.

Generative adversarial networks (GANs) are also a relatively new idea in machine learning\cite{goodfellow2014generative}. There have been many attempts at using GANs to improve existing models, such as a recent paper building on top of AudioUNet \cite{DBLP:journals/corr/abs-1903-09027}. GANs generally consist of two networks, one that generates samples and a new network whose goal is to discriminate between samples taken from the actual and generated distribution. This new network encourages the generator to produce more realistic samples than when trained without them. They are, however, notoriously difficult to train, particularly on one-dimensional data.


There have also been many recent advances in other, significantly more widely researched, super-resolution fields, such as image super-resolution. An early example of the work done for using deep learning to perform image super-resolution was produced in 2014 and made use of CNNs \cite{DBLP:journals/corr/DongLHT15}. This paper gives a good insight into the subject and involves techniques such as \textbf{patch extraction} to allow training over smaller datasets. Another example of recent innovation in this field is the MetaSRGan \cite{tan2020arbitrary}. This paper was adapted from a commonly used deep learning model to work on arbitrary scales by learning how to generate the weights of an up-sampling network instead of just learning the weights for a single scale.
\section{Contributions}
Our work builds upon previous super-resolution architectures and shows that a pre-upsampling model can be built to outperform existing approaches. Our pre-upsampling approach is approximately six times lighter weight and more effective at certain scales. This demonstrates the viability of a pre-upsampling approach and sets the groundwork for a new direction to take audio super-resolution research. Along with this new approach to audio upsampling, this paper explores the effects of adversarial learning on the more traditional pre upsampling architectures. We show that the presence of artifacts in a generators output makes it harder for a discriminator to produce an effective loss for training and propose solutions to the artifacting problem.

This paper also explores whether L1 or L2 loss leads to the best results on AudioEDSR and AudioUNET models.

All of the models explored in this paper are conceptually simple to understand and result in fully feedforward convolutional neural networks. This results in the models being fast to run and able to be run in parallel.

\section{Model Architectures}
As mentioned previously, this paper will go over three deep learning models trained for audio super resolution. This section will explain each in detail, including the motivations behind each approach and diagrams to make them easy to reproduce. The training algorithms have also been included in \Cref{ape:train}. A short summary of our models along with some key information about them is described in \Cref{tab:Sumaryintro}.
\begin{table}[H]
  \caption{A summary of key points for the different architectures}
  \label{tab:Sumaryintro}
\newcolumntype{K}{>{\centering\arraybackslash}X}
\begin{tabularx}{\textwidth}{K|K|K|KK|K|KK|K|K}
\toprule 
  & \multicolumn{3}{c}{AudioUNet} & \multicolumn{3}{c}{AudioEDSR} & \multicolumn{3}{c}{AudioUNetGAN} \\
\midrule 
 Pre/Post upsampling & \multicolumn{3}{c}{Pre-upsampling} & \multicolumn{3}{c}{Post-upsampling} & \multicolumn{3}{c}{Pre-upsampling} \\
Compatable scales & \multicolumn{3}{c}{All} & \multicolumn{3}{c}{Powers of 2} & \multicolumn{3}{c}{All} \\
Key Concept & \multicolumn{3}{c}{Bottle neck structure} & \multicolumn{3}{c}{Residual learning} & \multicolumn{3}{c}{Wasserstein GAN} \\
Important Citation(s) & \multicolumn{3}{c}{\cite{kuleshov2017audio,ronneberger2015unet,shi2016realtime,orhan2018skip}} & \multicolumn{3}{c}{\cite{lim2017enhanced,orhan2018skip,he2015deep}} & \multicolumn{3}{c}{\cite{DBLP:journals/corr/GulrajaniAADC17,arjovsky2017wasserstein,DBLP:journals/corr/LedigTHCATTWS16}} \\
 \bottomrule
\end{tabularx}

\end{table}
\newpage
\subsection{AudioEDSR}
\begin{figure}[th]
\centering
  \includegraphics[scale = 0.15]{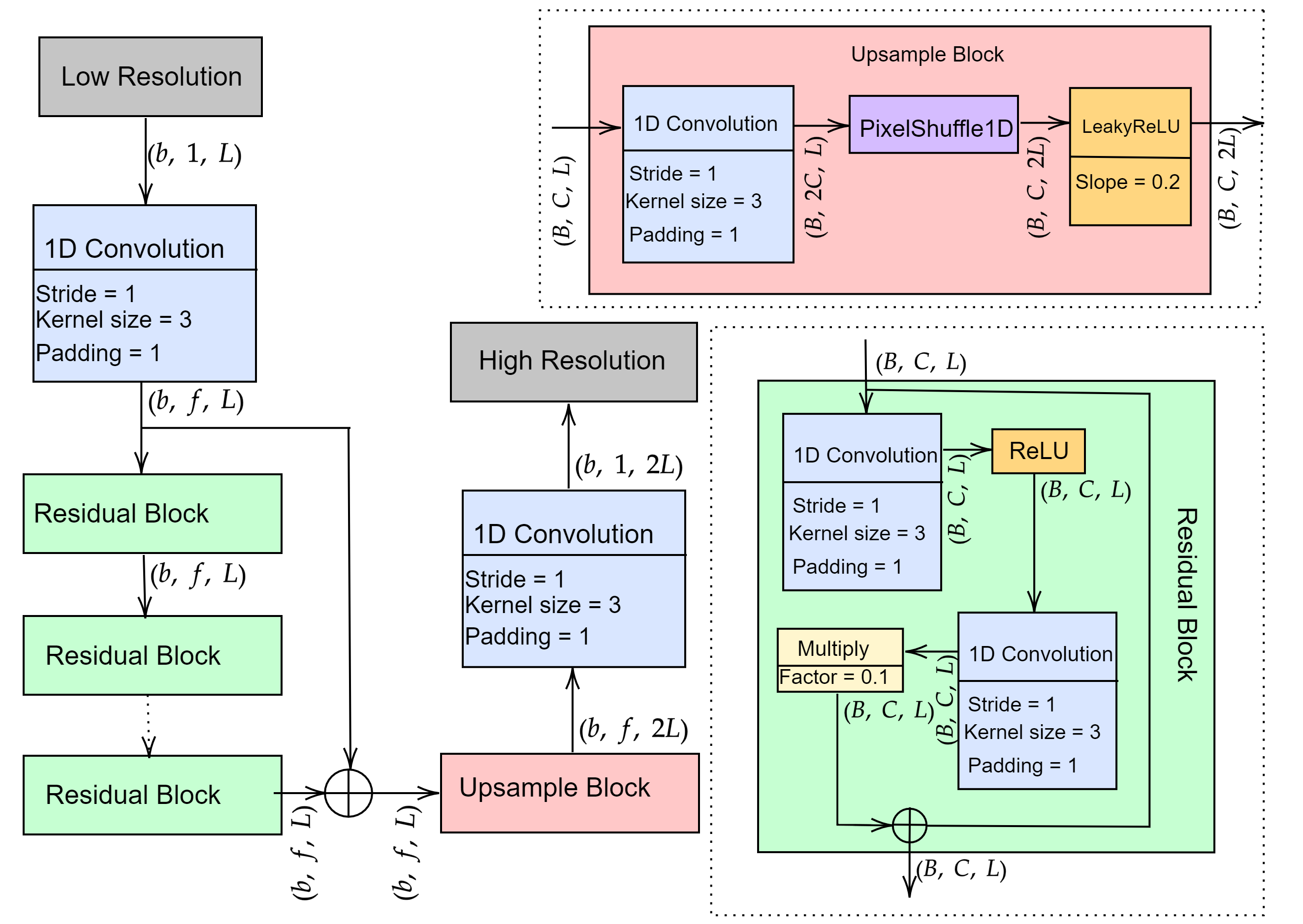}
  \caption{The structure of AudioEDSR network. The arrows show how the network layers are connected and are each labelled with the tensor shape they carry. The residual blocks all contain residuals in them, and the number $n$ of these blocks is a hyperparameter. The circle with a plus in it represents element-wise addition. $b$ is the batch size of an iteration, $L$ is the length of the signal, and $f$ is another hyperparameter representing the number of filters extracted. This model is for a two times upsampling problem. For the $2^q$ upsampling problem, the upsampling block is replaced by $q$ upsampling blocks in sequence.}
  \label{fig:EDSRNetwork}
\end{figure}

A popular type of deep Convolutional Neural Network, the Enhanced Deep Super-Resolution (EDSR) network was developed in mid-2017 \cite{lim2017enhanced} and was designed for image super-resolution. While many modern models tend to outperform the EDSR framework, it stood as a breakthrough in the residual super-resolution field. Its structures effective use of deep residual blocks for extracting features is not restricted to the image domain, and the design of the EDSR was adapted for application to audio in this paper. The adapted network is referred to as AudioEDSR throughout this report, and \Cref{fig:EDSRNetwork} shows its structure.

AudioEDSR is the post-upsampling model explored in this paper. This means that the Deep Neural Network performs alterations to the signal \textbf{before} standard methods perform upsampling. The motivation for this approach centres around minimising the number of parameters needed to train the model. This results in a faster model to train, and less memory is needed to store a pre-trained model. Both of these properties lead to less energy being expended altogether (currently a major problem facing deep learning).

The layered residual blocks used in this network are particularly effective in terms of feature extraction, with the network being capable of extracting more and more complex features in each layer. Our model also uses skip connections that help stabilise training, which has been demonstrated in the image equivalent of this model. The idea behind skip connections is to provide further context in learning new features so that the current layer has some information about the previous layer's input and output. This results in a more accurate model without excessive calculation overhead required if all previous layers were taken into account.
\newpage
\subsection{AudioUNet}
Another type of deep residual network was developed in 2015 and was named U-NET after the model's shape. The U-NET was originally developed for use in biomedical image segmentation \cite{ronneberger2015unet}. However, in 2017 Volodymyr Kuleshov \textit{et al} \cite{kuleshov2017audio} adapted this model's structure for use in an audio super-resolution task. This is probably the most state-of-the-art model to look at since it is still the one to beat, and there have been few improvements to its structure in the past four years. Although the authors have worked on subsequent models \cite{birnbaum2021temporal}, they are not fully feedforward, losing some training benefits. The model introduced by Kuleshov
 was implemented and is referred to it as AudioUNet in this document.
\begin{figure}[H]
\centering
   \includegraphics[scale = 0.16]{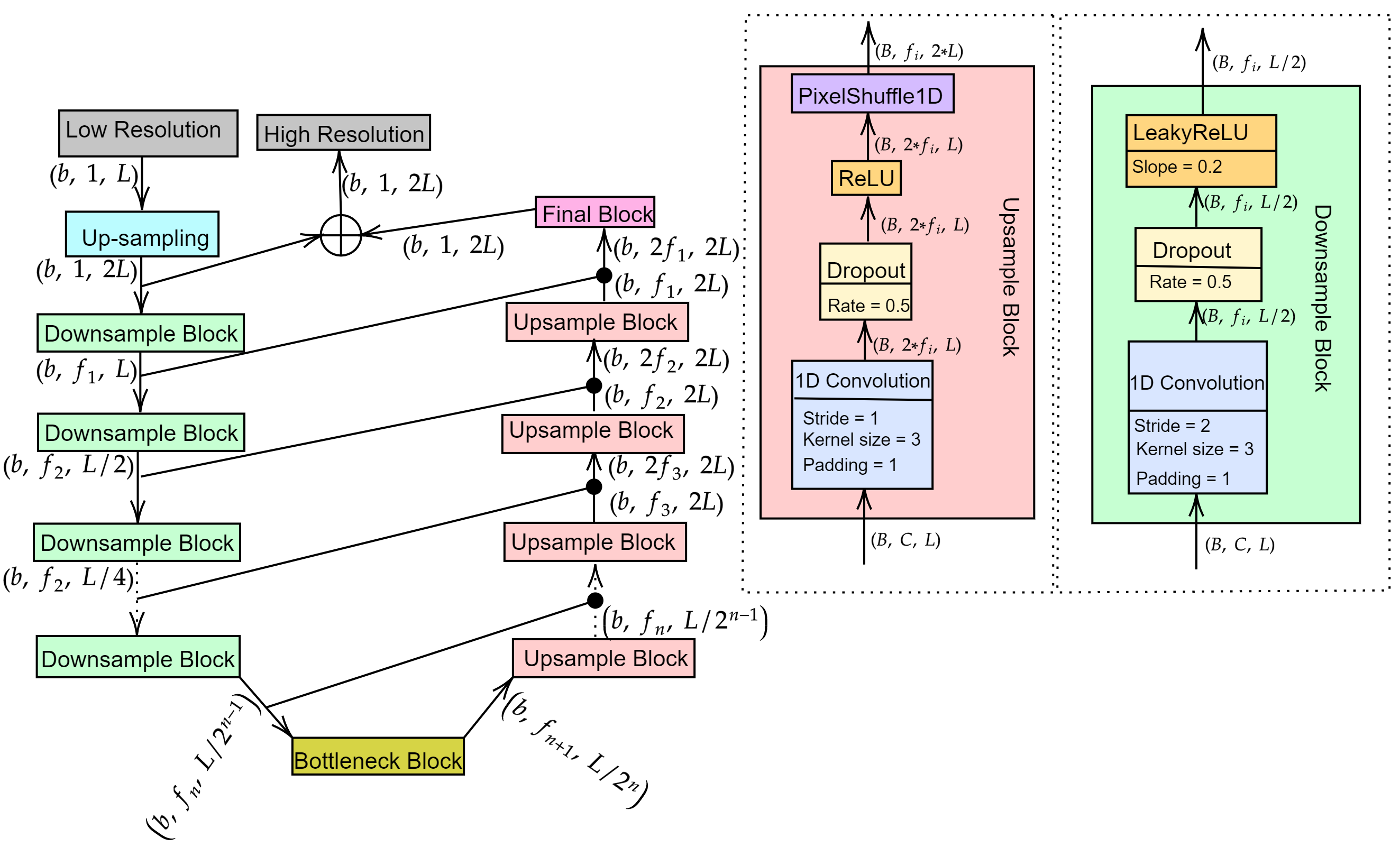}
  \caption{The structure of AudioUNet. Features $f_i$ are hyper-parameters, and the dark circles are stack connections that combine the features. $b$ is the batch size of an iteration, $L$ is the length of the signal, $n$ is the number of residual blocks.}
  \label{fig:UNETNetwork}
\end{figure}
AudioUNets structure works on blocks that downsample the input condensing the information further and further. At the lowest levels, the information stored in each block represents some abstract feature of the model. The signal is then upsampled using a 1D subpixel upsample layer\footnote{adapted from \cite{shi2016realtime}} and append the output of the result from the corresponding downsample block\footnote{cropped to be the same size as the upsample layer if needed}. This is achieved via skip connections. Skip connections have been shown to drastically improve training when the input distribution is highly correlated to the output \cite{orhan2018skip}. This is because stacking gives the upsample block more context of the detected features and should be present at these levels.

AudioUNet is the pre-upsampling model explored in this paper. This means that the Deep Neural Network performs alterations to the signal \textbf{after} standard methods perform upsampling. 

\newpage

\subsection{AudioUNetGAN}
AudioUNetGAN consists of a generative network that learns a distribution of outputs and a discriminator network
that calculates how believable the signal is. The SRGAN \cite{DBLP:journals/corr/LedigTHCATTWS16}
was the first to incorporate the idea of using a GAN in the super-resolution field and used a slightly different approach. This paper introduces a generator that is conditional on the low-resolution input data.  For this, it is no longer necessary to learn a distribution of outputs if only one deterministic output for each input is required. This means an input involving a random latent space is no longer needed. The model could be changed to learn a distribution of mappings for each input. It could also be changed to add a small latent space, but this is unnecessary for a functional model and would take longer to train for arguably limited benefit. This will be left as a potential route to explore in future development.
\begin{figure}[hbt]
    \centering
    \includegraphics[scale = 0.125]{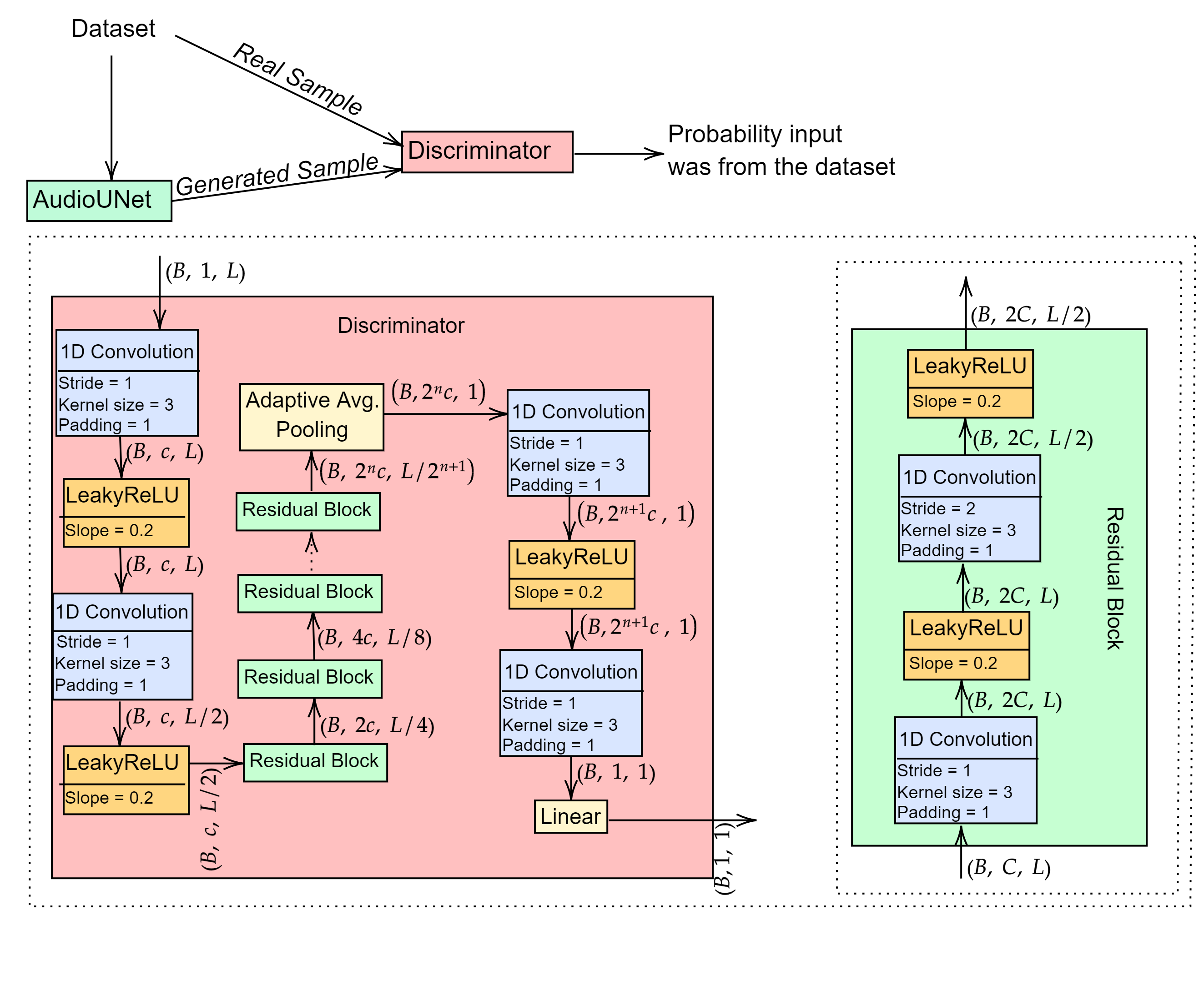}
    \caption{The training framework of AudioUNetGAN. The discriminator outputs a score for each input signal. $(B,1,L)$ represents the abstract shape of a tensor input to the block.}
    \label{fig:DiscrimModel}
\end{figure}

AudioUNetGAN will be a combination of the SRGAN and an improved GAN model called a Wasserstein-GAN (WGAN) \cite{arjovsky2017wasserstein}. The WGAN was designed in early 2017 and was a new approach to helping assisted generative modelling. This new model was designed to fix two main problems with the traditional GAN:\begin{itemize}[noitemsep,topsep=0pt]
    \item Updating two models simultaneously does not guarantee convergence, and the loss function's vanishing gradient is common. Finding the right hyper-parameters to put training into equilibrium is time-consuming.
    \item Traditional GANs are prone to mode collapse. This is when the generator learns how to trick the discriminator with a small range of generations and only outputs those regardless of input.
\end{itemize}
Fixing these two problems is ideal for any model so that AudioUNetGAN will make use of it.

Since the discriminator network no longer classifies the generated samples but scores how close they are, the literature on Wasserstein-based GANs often renames the discriminator to the `critic'. However, it serves mainly the same purpose in the network, and we shall continue referring to this sub-network as the discriminator.

\newpage
\section{Experiments}

\subsection{Dataset}
This paper uses CSTR VCTK corpus \cite{VCTK} as the primary dataset. This dataset contains speech from 110 different English speakers, each of whom read about 400 sentences from various newspapers. These sentences were chosen by an algorithm designed to maximise the phonetic coverage. All the samples in this dataset were recorded with the same microphones. The microphone has a sampling frequency of 96 kHz and a bit depth of 24 bits. All of the stored recordings were converted into having a depth of 16-bits and were downsampled to a 48 kHz sampling frequency.

The VCTK corpus was used because data transmitted is often vocal in communications. Making the data set specific to one type of audio will highlight the potential of this methodology since there will be more features to learn on. This dataset is used in other audio super-resolution works \cite{DBLP:journals/corr/abs-1903-09027, kuleshov2017audio} so it will be simple to benchmark the algorithm against existing baselines for evaluation.

Another thing to note is that artifacts may be introduced due to higher frequencies when downsampling a signal. These artifacts can act as a way of encoding information about which higher frequencies are present. So any model produced would be able to use this to its advantage to improve the accuracy of its predictions.  However, downsampling in this way does raise an interesting question about the problem statement. When a traditional condenser microphone records audio, it passes the signal through a band-pass filter. This removes detected frequencies that are not in the range their digitised form can represent without artifacting. In order to stay true to the problem statement, our models will upsample audio as though it had been recorded by a microphone and digitised to a lower sample rate. So when a signal is downsampled, it will be passed through an order eight low-pass Butterworth filter to remove the frequencies that the downsampled rate could not represent. This encoding of data via artifacting is something to consider for audio compression.
\subsection{Metrics for evaluation}
\label{sec:Evalprep}
\subsubsection{Signal-to-Noise Ratio}
Perhaps the most common method of evaluating audio quality is to measure its signal-to-noise ratio. The signal-to-noise ratio function compares a generated signal to the objective truth. It then calculates the proportion of the generated signal that is useful information (in Decibels). In this paper, the high-resolution audio will be considered the objective truth, and it will be compared to generated samples of the evaluated experiments.
\begin{equation}
\text{SNR}(x_{\text{generated}},x_{\text{actual}}) = 10\log_{10}\frac{||x_{\text{actual}}||^2_2}{||x_{\text{generated}}-x_{\text{actual}}||^2_2}
\end{equation}
A potential problem with using the original signal as the objective truth in SNR calculations is that these original signals will have at least some noise. This means that a better quality audio signal than the original may be produced, but it would produce a worse SNR than a slightly noisy signal.
\subsubsection{Log-Spectral Distance}
 The Log-Spectral Distance (LSD) was a metric analysed in the 1970s by Gray and Markel \cite{Gray1976DistanceMF} and was regarded as a way to give a better indication of the overall quality of a signal since this was not always reflected well in SNR. It instead computes how much distortion is present in the spectra of a signal. Again, this metric needs a ``true'' signal for comparison.
\begin{equation}
\text{LSD}(x_{\text{generated}},x_{\text{actual}}) = \frac{1}{W}\sum_{w=1}^W\sqrt{\frac{1}{K}\sum_{k=1}^n\Big(\log_{10}\frac{|x_{\text{generated}}(w,k)|^2}{|x_{\text{actual}}(w,k)|^2}\Big)^2}
\label{Equation:LSD}
\end{equation}
Where $w\in W$ represents the time windows and $k\in K$ represents the frequency bins. $x_{\text{generated}}$ represents the generated signal and $x_{\text{actual}}$ represents the expected output.

In their paper, Gray and Markel's suggested that the LSD metric is the best reference point for comparing speech processing methods.
\begin{figure}[H]
\vspace*{-0.5cm}
    \centering
    \includegraphics[width=.75\textwidth]{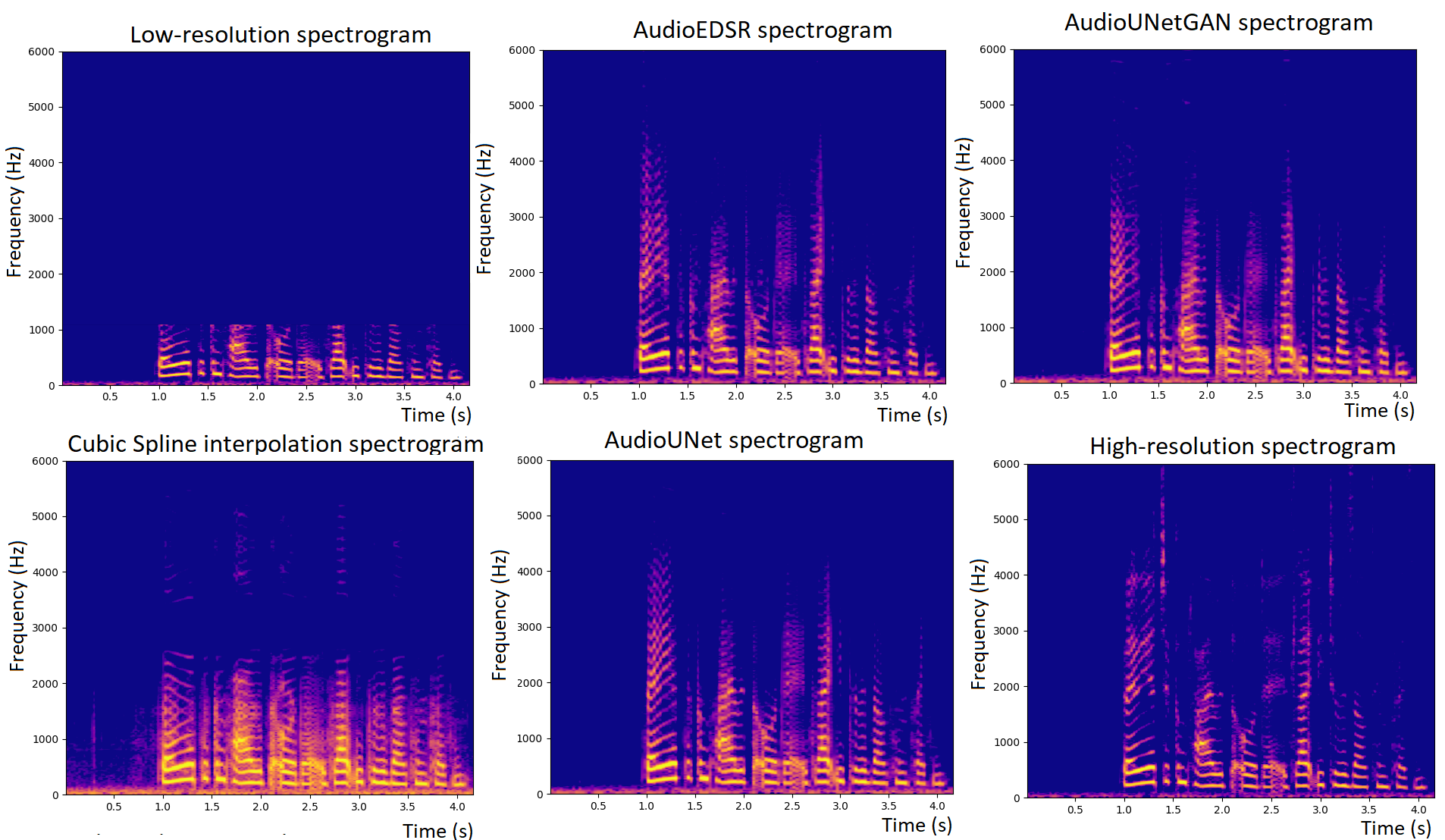}
    \caption{A visualisation of the performance of AudioEDSR, AudioUNet and AudioUNetGAN. compared to low-resolution input and the spline baseline.}
    \label{fig:AIRES}
\end{figure}
\subsubsection{MOS}
The results will also be evaluated using a mean opinion score (MOS). For this a group of people were asked to rate the overall quality of the sound produced by our models. MOS is used because it is a subjective metric, not only is it required for the signal to be accurate, but also to sound good, something this metric measures. In our experiments, people were asked to rate the quality on a scale of one to five, and then the arithmetic mean of their answers will be taken as the overall score. The tests were done anonymously\footnote{Methods were approved by institutions  ethics board}. The MOS metric is needed since a model may score well on the LSD and SNR metrics but produces audio that sounds clunky and robotic.
\subsubsection{L1 vs L2 loss}
Both AudioEDSR and AudioUNet can be implemented as stand-alone networks. The two most popular loss functions for the training of super-resolution models are L1 and L2 loss. AudioUNet paper \cite{kuleshov2017audio} uses L2 loss in training. This section investigates whether or not that was the optimal choice.
 
It has been shown that L1 can be superior to L2 for image processing when looking at the commonly chosen evaluation functions \cite{zhao2018loss}. This is believed to be because L2 loss has the property that it penalises outliers more, so can learn to train faster over data with minimal noise but when noise is present L1 can be a better choice.

To find out which loss function to use for AudioEDSR and AudioUNet, an investigation was carried out over the four times upsampling problem. The LSD and SNR results were compared for both models, and the results are shown in \autoref{tab:l1vl2}. Both models were trained over 32,100 batches which took approximately 7 hours each. The validation set was then used for evaluation, and the results show us that for AudioUNet, L1 loss looks like the best loss function to use since it scores more highly on both SNR and LSD. However, in AudioEDSR, the results show that SNR is better for L1 loss, but the LSD is better for L2 loss. AudioEDSR will be trained with L2 loss since, as previously discussed, LSD is generally a better metric, and the SNR score is within the bounds of error.

\begin{table}[H]
\centering
 \caption{\label{tab:l1vl2}A comparison of L1 loss vs L2 loss on the four times upsampling problem. The best results for each metric are in bold.}
 \newcolumntype{C}[1]{>{\centering\arraybackslash}p{#1}}
\begin{tabular}{C{1cm}|C{2.5cm}C{2.5cm}||C{2.5cm}C{2.5cm}}
\toprule 
  & \multicolumn{2}{c||}{AudioUNet} & \multicolumn{2}{c}{AudioEDSR} \\
\cmidrule{2-5} 
  & L1 & L2 & L1 & L2 \\
\cline{2-5} 
 SNR & $\bf{18.0 \pm 1.81}$ & $17.2\pm 1.66$ & $\bf{17.8\pm 1.96}$ & $16.4\pm 1.66$  \\
LSD & $\bf{2.20\pm 0.02}$ & $2.35\pm 0.02$ & $1.76 \pm 0.02$ & $\bf{1.60\pm 0.02}$  \\
 \bottomrule
\end{tabular}
\end{table}

\begin{table}[H]
\centering
    \caption{\label{tab:MLEVAL}A table showing the LSD, SNR and MOS scores of the spline and deep learning-based approaches. Values for LSD and SNR are in dB, and the MOS score was out of 5. The standard deviation is used as the bounds of error. The best performing metric for each scale is in bold.}
\newcolumntype{C}[1]{>{\centering\arraybackslash}p{#1}}
\begin{tabular}{C{2.35cm}|C{2.35cm}C{2.35cm}C{2.35cm}C{2.35cm}}
\toprule 
  & Spline & AudioEDSR & AudioUNet & AudioUNetGAN \\
\midrule 
 \multicolumn{5}{c}{LSD scores} \\
\midrule 
 2x Upsampling  & $\displaystyle 1.89\pm 0.003$ & $\displaystyle \mathbf{1.31\pm 0.009}$ & $\displaystyle 1.58\pm 0.009$ & $\displaystyle 1.52\pm 0.005$ \\
3x Upsampling  & $\displaystyle 2.95\pm 0.008$ & $\displaystyle N/A$ & $\displaystyle \mathbf{2.00\pm 0.01} 3$ & $\displaystyle 2.04\pm 0.001$ \\
4x Upsampling  & $\displaystyle 3.64\pm 0.011$ & $\displaystyle \mathbf{1.79\pm 0.023}$ & $\displaystyle 2.22\pm 0.019$ & $\displaystyle 2.11\pm 0.014$ \\
\midrule 
 \multicolumn{5}{c}{SNR scores} \\
\midrule 
 \multicolumn{1}{c|}{2x Upsampling } & $\displaystyle 2.90\pm 2.38$ & $\displaystyle \mathbf{23.9\pm 6.74}$ & $\displaystyle 22.1\pm 2.10$ & $\displaystyle 20.8\pm 1.53$ \\
3x Upsampling  & $\displaystyle -0.37\pm 1.99$ & $\displaystyle N/A$ & $\displaystyle \mathbf{19.1\pm 1.44}$ & $\displaystyle 17.8\pm 1.44$ \\
4x Upsampling  & $\displaystyle -1.94\pm 1.24$ & $\displaystyle 17.6\pm 1.78$ & $\displaystyle \mathbf{18.3\pm 1.93}$ & $\displaystyle 16.9\pm 1.37$ \\
\midrule 
 \multicolumn{5}{c}{MOS scores} \\
\midrule 
 2x Upsampling  & $\displaystyle 3.80\pm 1.067$ & $\displaystyle \mathbf{4.06\pm 0.809}$ & $\displaystyle 3.82\pm 0.529$ & $\displaystyle 3.94\pm 0.559$ \\
3x Upsampling  & $\displaystyle 2.82\pm 0.404$ & $\displaystyle N/A$ & $\displaystyle \mathbf{3.47\pm 0.390}$ & $\displaystyle 3.18\pm 0.779$ \\
4x Upsampling  & $\displaystyle 1.94\pm 0.434$ & $\displaystyle 2.24\pm 0.191$ & $\displaystyle \mathbf{2.76\pm 0.691}$ & $\displaystyle 2.06\pm 0.559$ \\
 \bottomrule
\end{tabular}
\end{table}
\section{Results}
\begin{wraptable}{r}{6cm}
 \vspace*{-0.8cm}
     \caption{\label{tab:PARAMETERS}A table showing the number of parameters AudioEDSR and AudioUNet models contain.}
 \newcolumntype{C}[1]{>{\centering\arraybackslash}p{#1}}
\begin{tabular}{C{1cm}|C{2cm}C{2cm}}
\toprule 
  & AudioEDSR & AudioUNet \\
\midrule 
 2x & $\displaystyle 9,594,113$ & $\displaystyle 56,411,394$ \\
4x & $\displaystyle 9,692,673$ & $\displaystyle 56,411,394$ \\
 \bottomrule
\end{tabular}

\end{wraptable}
As seen in \autoref{tab:MLEVAL} AudioEDSR model produces better results for the two times upsampling case and performs only marginally worse on the four times upsampling case. Our results also show that all of the deep learning models outperformed the spline approach in every metric. This showcases the effectiveness of deep learning on audio super-resolution tasks and that the EDSR and GAN can be successfully adapted to the audio domain.

Some of the results are surprising, though, especially when comparing the results from the three deep learning models. For example, the post-upsampling AudioEDSR model performed best on all metrics in the two times upsampling case. This was unexpected since all recent models that perform audio super-resolution have been based on the pre-upsampling framework. A benefit to using AudioEDSR network is that it uses significantly fewer parameters to train, meaning it is more lightweight, so it takes up less space in memory and is faster to train. This difference can be seen in \autoref{tab:PARAMETERS} where it is clear that AudioEDSR is approximately $83\%$ smaller than AudioUNet.

Despite its performance in the two times upsampling case, AudioEDSR falls short of AudioUNet's MOS in the four times upsampling problem. This is likely due to the tonal artifacting added by the subpixel shuffle layer, making it harder for this model to train and it even audible to some listeners. Although AudioUNet also suffers from this type of artifacting, its structure of skip connections makes it easier for the model to learn to lessen the effect. Notably, though, AudioEDSR still outperforms AudioUNet on the LSD metric and is not that far behind on the SNR front, meaning the lightweight nature may still make it a better option for real-world deployment.

Another surprise is that the GAN does not provide much improvement over the original purely L1 based training of AudioUNet, performing slightly worse in almost all metrics. The one metric that it beats the standard AudioUNet in is the MOS on the two times upsampling problem. The leading theory for this lack of improvement is that the systematic tonal artifacts produced by AudioUNet, are predominately used in the discriminators scoring process. This is backed up by almost all of AudioUNetGANs performance metrics getting further from AudioUNets as the upsampling factor increases since the intensity of these tonal artifacts increases. The next section explores this artifacting and explains why it is a problem when training a GAN.

\begin{figure}[tbh]
 \hspace*{-0.5cm}
    \centering
\makebox[\textwidth][c]{\includegraphics[width=\textwidth]{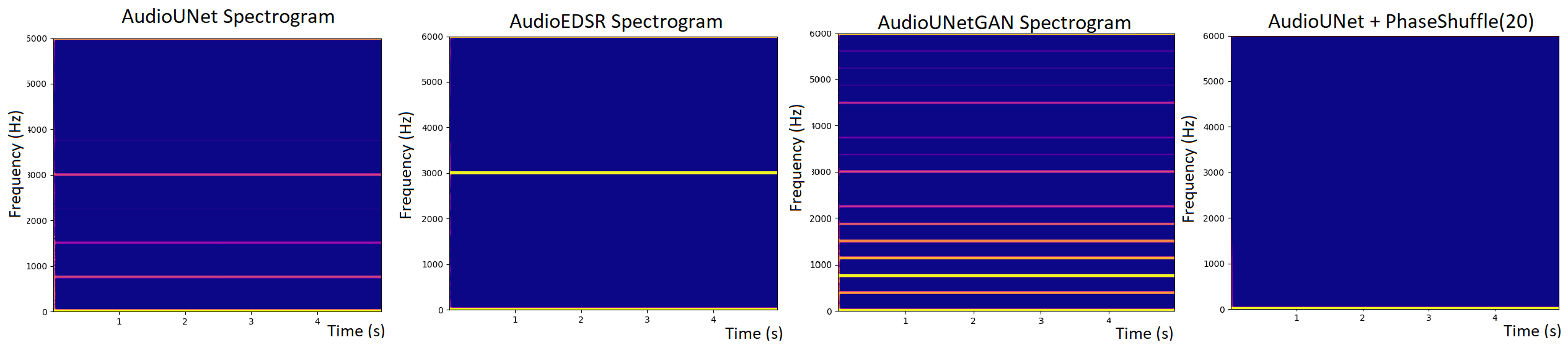}}
    \caption{Spectrograms showing the tonal artifacts produced by the four times upsampling layer of the three Deep learning models when the input is an empty signal. The image also shows how modifying the network with phase shuffles can reduce these artifacts.}
    \label{fig:artifactsnoshuffle}
\end{figure}

\section{Limitations of Approach}
\subsection{Artifacts}

One of the major drawbacks of our approach is the spectral arifacts that are added. The main form of artifacting present in our approach are tonal artifacts. A tonal artifact is an unwanted additional sound with constant frequency (a tone). These can be introduced by using overlapping deconvolution layers for a similar reason to the 2D checkerboard example. The model should be able to learn how to correct these issues. However, because the model is fully convolutional and the signal may not have a length equal to multiple stride, there will likely not be a consistent degree of overlap at each stage. Even when there is constant overlap throughout the signal, there is not at the sides of the signal. This is called boundary artifacting.

To avoid these tonal artifacts being added due to overlapping, a one-dimensional version of the sub-pixel convolution network, as described by Wenzhe Shi in a recent paper \cite{DBLP:journals/corr/ShiCHTABRW16}, can be used. This sub-pixel convolution uses a deep convolutional neural network to increase the number of channels the model is currently operating on by the upsample rate. Then the sub-pixel shuffle is used to reorder these into the feature space. This means that only a whole number upsample is possible, but a recent work \cite{tan2020arbitrary} shows an alternative dimension shuffle layer can be used to remove this condition.


Other ways tonal artifacts can be introduced include poor weight initialisation or loss functions. This would be a particular problem for a model that uses convolution filters since they are used repeatedly in series. Poor choices for initialised weights can introduce unwanted frequencies. Loss functions can cause high-frequency tonal artifacts in the gradients generated by backpropagation. It has been observed that it is an issue particularly in audio-based models \cite{pons2021upsampling}, and it is thought to be caused by higher frequency components being present in audio. It is also a significant problem in adversarial learning, and paper \cite{DBLP:journals/corr/abs-1802-04208} finds that these loss based tonal artifacts occur at predictable phases, making it easy for an adversary to discern fake samples. A future area of advancement to this approach could be altering the discriminator's kernels phase and adding more random noise into the generator at certain intervals.

\section{Conclusion}
This paper has demonstrated that deep learning can be an effective tool in audio super-resolution tasks. Post-upsampling models were shown to be a competitive alternative to the current pre-upsampling state of the art. It is also evident from our investigations that a GAN framework could also be a useful tool in this domain, but further research needs to be done to unlock its true benefits.

Our current models perform a deterministic mapping from the low-resolution signal to the super-resolved signal. The model could be changed to learn a distribution by adding some noise into the input. Alternatively stochastic differential equations (SDEs) could be used to train AudioUNetGAN. SDEs are a recent development to the field \cite{SDE} and are seeing a lot of recent research \cite{SDEREV}.

Seeing how the model performs on different domains would also be of interest to me and a potential investigation for the future. Some examples of other domains you could apply this model to are: Recordings of different animals, recordings of the sea and music.
\newpage
\bibliographystyle{plain}
\bibliography{references}
\newpage
\appendix
\section{Training algorithms\label{ape:train}}

\begin{algorithm}[H]
\caption{AudioEDSR training}
\begin{algorithmic}

\STATE Use hyper-parameters $\alpha = 0.0001, \beta_1 = 0.9,\beta_2 = 0.999$
\FOR {K steps}
    \STATE {Sample high resolution data $x = \Tilde{w}(t;S)$ from current batch}
    \STATE {Generate $l = \Tilde{w}(t;\hat{S})$ from $x = \Tilde{w}(t;S)$}
    \STATE {$\Tilde{x} \gets G_{\theta}(l)$}
    \STATE {$L\gets     \frac{1}{n}\sum_{i=1}^n|\Tilde{x}-x|^2$}
    \STATE $\theta \gets \text{Adam}(\nabla_{\phi}\frac{1}{m}\sum_{i=1}^mL,\theta,\alpha,\beta_1,\beta_2)$
\ENDFOR
\end{algorithmic}
\end{algorithm}
\begin{algorithm}[H]
\caption{AudioUNet training}
\begin{algorithmic}
\STATE Use hyper-parameters $\alpha = 0.0001, \beta_1 = 0.9,\beta_2 = 0.999$
\FOR {K steps}
    \STATE {Sample high resolution data $x = w_D(t;S)$ from current batch}
    \STATE {Generate $l = w_D(t;\hat{S})$ from $x = w_D(t;S)$}
    \STATE {$\Tilde{x} \gets G_{\theta}(l)$}
    \STATE {$L\gets     \frac{1}{n}\sum_{i=1}^n|\Tilde{x}-x|$}
    \STATE $\theta \gets \text{Adam}(\nabla_{\phi}\frac{1}{m}\sum_{i=1}^mL,\theta,\alpha,\beta_1,\beta_2)$
\ENDFOR
\end{algorithmic}
\end{algorithm}
\begin{algorithm}[H]
\caption{AudioUNetGAN training}
\begin{algorithmic}
\STATE Use hyper-parameters $\lambda =10, n_{\text{critic}} = 5, \alpha = 0.0001, \alpha = 0.0001, \beta_1 = 0.9, \beta_2 = 0.999$

\WHILE {$\theta$ not converged}
    \FOR {$t = 1,\hdots,n_{\text{critic}}$}
        \FOR {$i = 1,\hdots,m$}
            \STATE {Sample high resolution data $x = \Tilde{w}(t;S)$ from current batch}
            \STATE {Generate $l = \Tilde{w}(t;\hat{S})$ from $x = \Tilde{w}(t;S)$}
            \STATE {$\Tilde{x} \gets G_{\theta}(l)$}
            \STATE {$\hat{x} \gets \epsilon x + (1-\epsilon) \Tilde{x}$}
            \STATE {$L^{(i)}\gets D_{\phi}(\Tilde{x})-D_{\phi}+\lambda(||\nabla_{\hat{x}}D_{\phi}(\hat{x})||_2-1)$}
        \ENDFOR
        \STATE $\phi \gets \text{Adam}(\nabla_{\phi}\frac{1}{m}\sum_{i=1}^mL^{(i)},\phi,\alpha,\beta_1,\beta_2)$
    \ENDFOR
        \STATE {Generate a batch of m low resolution variables $l = \Tilde{w}(t;\hat{S})$ from $x = w_D(t;S)$}
        \STATE $\theta \gets \text{Adam}(\nabla_{\theta}\frac{1}{m}\sum_{i=1}^m[-D_{\phi}(G_\theta(l))],\theta,\alpha,\beta_1,\beta_2)$
\ENDWHILE
\end{algorithmic}
\end{algorithm}
\end{document}